\documentclass[useAMS,usenatbib]{mn2e}

\usepackage{epsfig,graphicx,times}

\newcommand{\kms}{{{km~s$^{-1}$}}}
\newcommand{\teff}{{$T_\mathrm{eff}$}}
\newcommand{\logg}{{log~{\em g}}}

\newcommand{\vt}{{$\xi_\mathrm{t}$}}
\newcommand{\vsini}{{\em v}\,sin\,{\em i}}

\newcommand{\aaps}{A\&A}
\newcommand{\aj}{AJ}
\newcommand{\mnras}{MNRAS}

\def\deg{\hbox{$^\circ$}}

\title[The iron abundance of the Magellanic Bridge]
{The iron abundance of the Magellanic Bridge
\thanks{Based on observations made with the NASA/ESA Hubble Space 
Telescope, obtained at the Space Telescope Science 
Institute, which is operated by the Association of Universities 
for Research in Astronomy, Inc., under NASA contract NAS 5-26555. 
These observations are associated with programs 6073, 8161 
and 9130.}}
\author[P.L. Dufton, R.S.I. Ryans, H.M.A. Thompson and R.A. Street]
{P.L. Dufton$^{1}$\thanks{E-mail:p.dufton@qub.ac.uk},
R.S.I. Ryans$^{1}$, H.M.A. Thompson$^{1}$ and R.A. Street$^{2,3}$\\
$^{1}$Department of Physics and Astronomy, Queen's University of Belfast,
Belfast BT7 1NN, United Kingdom.
\\
$^{2}$ Las Cumbres Observatory, 6740B Cortona Drive, Goleta, CA 93117, USA.
\\
$^{3}$ Department of Physics, Broida Hall, University of California, 
Santa Barbara, CA 93106-9530, USA.}
\begin{document}

\date{Accepted  Received ; in original form}

\pagerange{\pageref{firstpage}--\pageref{lastpage}} \pubyear{}

\maketitle

\label{firstpage}

\begin{abstract}

High-resolution HST ultra-violet spectra for five B-type stars  in the
Magellanic Bridge and in the Large and Small Magellanic Clouds have
been analysed to estimate their iron abundances. Those for the Clouds
are lower than estimates obtained from late-type stars or the optical
lines in B-type stars by approximately 0.5 dex. This may be due 
to systematic errors possibly
arising from non-LTE effects or from errors in the atomic data as
similar low Fe abundances having previously been reported from the
analysis of the ultra-violet spectra of Galactic early-type stars. The
iron abundance estimates for all three Bridge targets appear to be
significantly lower  than those found for the SMC and LMC by
approximately -0.5 dex and -0.8 dex  respectively and these 
differential results should not be affected by any 
systematic errors present in the absolute abundance estimates.
These differential iron abundance
estimates are consistent
with the underabundances for C, N, O, Mg  and Si of approximately -1.1
dex relative to our Galaxy previously found in our Bridge targets. The
implications of these very low metal abundances for the Magellanic
Bridge are discussed in terms of metal deficient material being
stripped from the SMC.

\end{abstract}

\begin{keywords}

Magellanic Clouds --  galaxies: interactions -- galaxies: abundances --
stars: abundances -- stars: early-type

\end{keywords}

\section{Introduction}

The Large (LMC) and Small (SMC) Magellanic Clouds are irregularly 
shaped galaxies, which are gravitationally bound to our Milky Way. In
addition to these stellar systems, there are two regions of neutral
hydrogen, designated  the Magellanic `Stream' and `Bridge'. The former
traces out a large  arc in the sky from between the Clouds to beyond
the South Galactic Pole  (Fujimoto \&\ Murai \citealp{Fuj84}). The
latter is a band of  material which joins the two Clouds (Hindman et
al. \citealp{Hin63}; Staveley-Smith et al. \citealp{Sta98}; Muller et
al. \citealp{Mul04})  and is widely considered to be a remnant of a
close tidal encounter  of the two Clouds about $\sim$200 Myr ago
(Murai and Fujimoto \citealp{Mur80}; Gardiner et al. \citealp{Gar94}).
This produced a tidal  bridge and tail structure (Gardiner \& Noguchi
\citealp{Gar96}; Sawa, Fujimoto \& Kumai \citealp{Saw99}), which is
viewed as  overlapping in the sky, with  the tail material modelled to
be more distant than the bridge and with a larger radial velocity. 

As H\,{\sc i} emission and stellar condensation are  strongly
correlated, it is natural to suppose that star-formation may  take
place in these regions. Additionally, molecular clouds have  been
identified in the Magellanic Bridge from CO emission (Mizuno et al.
\citealp{Miz06}), which given the short lifetimes of such clouds were
probably formed after the tidal encounter. Attempts to find a stellar
population  associated with the Stream have been largely
unsuccessful  (Irwin \citealp{Irw91}). By contrast, young stellar
associations have  been identified in the Bridge (Irwin et al.
\citealp{Irw85,Irw90};  Demers \&\ Irwin \citealp{DIr91}; Bica and
Schmitt \citealp{Bic95})  and may reflect an episode of star
formation triggered by the close encounter of the two Clouds. 

The chemical composition of both the SMC and LMC have been
extensively  studied using both stellar targets and emission nebulae.
For example the  spectra of early-type stars have been used, viz,  %
O-type supergiants (Crowther et al. \citealp{Cro02}; Hillier et al. 
\citealp{Hil03}),  O-type main sequence targets (Bouret et al.
\citealp{Bou03}), B-type supergiants (Trundle et al. \citealp{Tru04};
Dufton et al.  \citealp{Duf05}),  B-type giants (Korn et al.
\citealp{Kor00}; Lennon et al. \citealp{Len03}) B-type main sequence
stars (Korn et al. \citealp{Kor02, Kor05}, Hunter et al. \citealp{Hun05,Hun07})
and A-type supergiants (Venn \citealp{Ven99}; Venn \&
Pryzbilla  \citealp{Ven03}). Late-type stellar investigations have
included  FG-type supergiants (Russell \& Dopita \citealp{Rus92};
Spite et al.  \citealp{Spi89, Spi91}; Hill et al. \citealp{Hil95}; 
Andrievsky et al. \citealp{And01}), Cepheids (Luck et al.
\citealp{Luc98})  and K-type supergiants (Barbuy et al.
\citealp{Bar91};  Hill \citealp{Hil97, Hil99}; Gonzalez \& Wallerstein
\citealp{Gon99}). Additionally there have been many H {\sc ii} region
studies  (see, for example, Kurt et al. \citealp{Kur99}; Peimbert et
al.  \citealp{Pei00}; Testor \citealp{Tes01}; Tsamis et al.
\citealp{Tsa03}).  In general terms, these imply that the LMC and SMC
are underabundant  in metals (compared with our Galaxy) by
approximately -0.3 and -0.6 dex,  with nitrogen showing a greater
underabundance particularly for unevolved stellar targets and H {\sc
ii} regions.

The chemical composition of the Magellanic Bridge is more uncertain. 
Hambly et al. (\citealp{Ham94}) obtained optical spectra for two
early-type stars (DI~1194 and DI~1388) in order to investigate whether
their  chemical compositions reflected those of the LMC or SMC.
However,  it was only possible to make marginal identifications of a
few metal lines,  and these implied large but uncertain metal
depletions. Rolleston et al. (\citealp{Rol99}) analysed high and
medium  resolution spectroscopy of three other B-type bridge stars. 
They found an underabundance in the light metals (C, N, O, Mg, Si)  of
approximately -1.1 dex compared with our Galaxy. Additionally they 
undertook a differential analysis relative to the SMC B-type main
sequence star, AV~304, and found the Bridge material to be relatively
underabundant  in metals by -0.6 dex. Unfortunately no iron features
were observed in these optical spectra  thereby preventing an abundance
determination for this element. Lehner et al.\ (\citealp{Leh01})
have studied the  interstellar ultraviolet absorption and H {\sc i}
emission spectra  towards a young star in the Bridge. These
observations also were consistent with a low metallicity ($\simeq
-1.1$ dex) compared  to that of our Galaxy. Recently Lee et al.
(\citealp{Lee05}) studied B-type targets in the SMC south east wing
which adjoins the Bridge. These  were found to have a normal SMC
chemical composition, although a  re-analysis of the extant Bridge
observational using sophisticated non-LTE techniques confirmed the
very low metal abundances found previously by Rolleston et al.

The low metal abundance deduced for the Bridge material is surprising
as one would expect its metallicity to reflect that of its progenitor
material. A possible explanation is that the inter-cloud  material
could have been stripped off from either of the galaxies at a  much
earlier encounter (about $\sim$8 Gyr ago; Kobulnicky \& Skillman
\citealp{Kob97}; Da Costa \& Hatzidimitriou \citealp{DaC98}), when
less nucleo-synthetic processing of interstellar medium of the SMC had
occurred. However, this is inconsistent with the numerical
simulations of, for example, Murai and Fujimoto (\citealp{Mur80}) and
Gardiner and Noguchi (\citealp{Gar96}) that imply that the Magellanic
Bridge was formed  relatively recently and also with the presence of
early-type stars, implying recent  star formation. 

Here we present and analyse Hubble Space Telescope (HST) ultra-violet
spectroscopy of three Bridge B-type stars, together with similar
targets in the SMC and LMC. The  observational data cover the
wavelength range between approximately 1900   to 2000\AA, where
the spectrum is dominated by absorption lines due to iron group
elements. Hence the principle aim of this analysis is to investigate
whether the very low light metal abundances inferred for the  Bridge
material is replicated in the heavier metals.

\begin{center}
\begin{table*}
\caption[]{Details of the HST observations.}
\begin{tabular}{llcclccrccccc}\hline
Star   & Region    &     RA   & DEC       & Sp. Ty. & V &  E(B-V) & Inst. & Proposal 
& Wavelength & Exp. Time  & $v_{\rm LSR}$ & S/N 
\\
          & &\multicolumn{2}{c}{2000.0} & mag & mag &
& & & \AA & seconds & \kms & ratio
\\\hline
               
AV\,304      & SMC      & ~1 02 21.4 & -73 39 15 & B0.5V &14.98 & 0.03 & STIS  & 7759     
& 1888-1978  & 5940  & 145  & 30
\\
DGIK\,975    & Bridge   & ~4 19 58.6 & -73 52 26 & B2 & 15.05 & 0.12 & STIS  & 9130     
& 1888-1978  & 21075 & 190  & 29
\\
DI\,1239     & Bridge   & ~2 30 40.8 & -74 04 47 & B1 & 15.24 & 0.08 & STIS  & 9130     
& 1888-1978  & 21075 & 159  & 37
\\
DI\,1388     & Bridge   & ~2 57 11.9 & -72 52 55 & B0 &14.39 & 0.08 & GHRS  & 6073     
& 1888-1929  & 1410  & 140  & 18
\\
LH\,10-3270  & LMC   & ~5 57 21.2 & -66 25 01 & B1 V & 14.90 & 0.10 & STIS  & 9130     
& 1888-1978  & 20480 & 313  & 42
\\
\hline
\end{tabular}
\end{table*}
\label{Obs_data}
\end{center}

\section{Observations and data reduction} \label{Obs_red} 
HST spectroscopy of the Magellanic Bridge stars have been obtained
during two campaigns that utilised the GHRS and the STIS
spectrographs. The former obtained spectroscopy of a Bridge 
target, DI\,1388, in the  wavelength range 1888-1930\AA\ with 
a spectral resolution of approximately 14 \kms whilst the latter
observed two Bridge targets and the LMC star, LH\,10-3270 between 1888
and 1970\AA\ with a spectral resolution of approximately 15 \kms. 
These observations were supplemented with archive  STIS
observations of the SMC star, AV\,304, that had the same
wavelength region and spectral resolution. 
The observational data are summarized in Table
\ref{Obs_data} with the V-magnitudes and spectral types
being taken from Azzopardi et al.
(\citealp{Azz75}), Grondin et al. (\citealp{Gro90}), Demers \& Irwin 
(\citealp{DIr91}), Demers et al. (\citealp{Dem91}), Parker et al. 
(\citealp{Par92}), Garmany et al. (\citealp{Gar87}) and 
Rolleston et al. (\citealp{Rol99}).

For the STIS observations, multiple exposures each of approximately
3000 seconds were obtained. These were combined using the {\sc
scombine} procedure in {\sc iraf}, with tests indicating that the
actual choice of  parameters was not critical. Both the combined  STIS
and the GHRS spectra were input into the data reduction package {\sc
dipso} (Howarth et al.\ \citealp{How94}) and  cross-correlated with
theoretical spectra (see Sect. \ref{spec_anal} for details) to obtain
the stellar radial velocity. These are also listed in Table
\ref{Obs_data} and are in good agreement with those found previously
(Hambly et al. \citealp{Ham94}; Rolleston et al. \citealp{Rol03}),
apart from the LMC target, LH\,10-3270. Previously Rolleston et al.
(\citealp{Rol02}) had found a value of 197 \kms, which  is smaller
than the mean LMC radial velocity of 261\kms (Luks \&
Rohlfs (\citealp{Luk92}) and the estimate of 
approximately 310 \kms found here. Hence this star may be a single
lined spectroscopic binary. The spectra were corrected for these
radial  velocity shifts prior to further analysis.

It was difficult to estimate signal-to-noise (S/N) ratios due the rich 
aborption line spectra. We have used synthetic and the relatively 
narrow lined observational spectrum of AV\,304 to aid the 
identification of  suitable regions and these have been fitted by  low 
order polynomials to yield the S/N ratios listed in Table
\ref{Obs_data}.  Given that as it was effectively impossible to
identify true continuum regions, these values should be considered as
lower limits. These estimates can be compared with those from the STIS
and GHRS Exposure Time calculators which predicted S/N ratios of 25 for AV\,304
and 30 for all the other targets. These predicted values are very sensitive 
to the adopted reddennings, which may partially explain the differences with
those listed in Table \ref{Obs_data}.

\section{Data analysis} \label{spec_anal}

The analysis is based on grids of non-LTE model atmospheres that have
been calculated using the codes {\sc tlusty} and {\sc synspec} 
(Hubeny \citealp{Hub88}; Hubeny \& Lanz \citealp{Hub95}; Hubeny et 
al.\, \citealp{Hub98}).  Details of the methods can be found in Ryans
et al.\  (\citealp{Rya03}), while the grids have been discussed in
more  detail by Dufton et al.\ (\citealp{Duf05}). 

Briefly four grids have been generated with base metallicities
corresponding  to our galaxy ($[\frac{Fe}{H}]$ = 7.5 dex) and with
metallicities reduced by 0.3,  0.6 and 1.1 dex. These lower
metallicities were chosen so as to be representative  of the LMC, SMC
and Bridge material. For each base metallicity, models have been
calculated covering a range of effective temperature  from 12\,000 to
35\,000 K in steps of 1000K or less and logarithmic gravities  (in cm
s$^{-2}$) from 4.5 dex down to close to the  Eddington limit  (which
will depend on the effective temperature) in steps of 0.25 dex. For
each pair of effective temperatures and surface gravities, five models
with microturbulences (\vt) of 0, 5, 10, 20 and 30 \kms were generated. 

Atmospheric parameters (summarized in Table \ref{Atm_par}) for all 
our targets have been previously estimated from optical  spectra by
Rolleston et al. (\citealp{Rol99}) for the Bridge stars, Rolleston et
al. (\citealp{Rol02}) for LH\,10-3270 and Hunter et al.
(\citealp{Hun05}) for AV\,304. The first two analyses used an LTE
approach based on the model atmospheres of Kurucz (\citealp{Kur91}),
whilst the last was based on the non-LTE grid utilised here. However
atmospheric parameters for AV\,304 estimated by Rolleston et al. 
(\citealp{Rol03}) using Kurucz's LTE models are in excellent
agreement  with those of Hunter et al.\ and hence the adoption of LTE
estimates should be acceptable; we return to this point when
considering  possible sources of errors in our iron abundance
estimates.

\begin{center}
\begin{table}
\caption[]{Atmospheric parameters and projected rotational 
velocities adopted for our targets.}
\begin{tabular}{lccccc}\hline
Star         & $T_{\rm eff}$    & \logg       &   \vt     & \vsini  
\\
             & K                & cm s$^{-2}$  &  \kms    &  \kms 
\\\hline
AV\,304      & 27\,500          & 3.90         & 3        &  11  
\\              
DGIK\,975    & 20\,000          & 3.60         & 5        &  120 
\\              
DI\,1239     & 24\,000          & 3.80         & 5        &  90  
\\              
DI\,1388     & 32\,000          & 4.00         & 5        &  180 
\\
LH\,10-3270  & 29\,500          & 4.10         & 6        &  30  
\\
\hline
\label{Atm_par}
\end{tabular}
\end{table}
\end{center}

Using a grid model with an appropriate metallicity and atmospheric 
parameters closest to the values
listed in Table \ref{Atm_par}, we have synthesised the spectral region
observed by HST for each star. As the spacing between grids points
is smaller than the typical uncertainties in estimating atmospheric
parameters, the adoption of the nearest grid point  is unlikely to be
a significant source of error. For elements (carbon, nitrogen, oxygen,
magnesium and silicon) explicity included in the {\sc tlusty}
calculation, these calculations were in a non-LTE approximation. Iron
was also included in the model atmosphere calculations but due to the 
complexity of its Grotrian diagram, individual levels were combined 
together into superlevels. Hence for the spectral synthesis, iron and 
all other elements were treated in LTE. Test calculation showed that
typically 80\% of the absorption in this spectral region arose from
iron with over 95\% being due to iron group elements -- hence these
are effectively LTE calculations. To fit an observed spectrum,
calculations were  undertaken for the set of element abundances
adopted for a particular grid (see Dufton et al. \citealp{Duf05} for
details) and with  all these abundances scaled. The actual element
abundances for any given star may not vary in step but as iron
dominates the absorption, any differential variations in element
abundances should not significantly affect these theoretical spectra. 

This approach has the consequence of introducing an 
inconsistency between the abundances adopted in the model atmosphere 
and spectrum synthesis calculations. For example, for the star, 
AV\,304, a model would be adopted from the grid with a metallicity 
appropriate to the SMC viz. -0.6 dex that of the Galaxy. 
Spectra would then be synthesised for different metallicities
which might range, for example, from the Galactic composition 
to -1.2 dex of the Galactic composition. We have investigated
the consequences of this simplification as folows. Firstly
we calculated the spectrum for an SMC model of AV\,304 with 
atmospheric parameters, \teff=27\,500, \logg =3.9, \vt=5 \kms 
and with a metallicity -0.6 dex that of the Galaxy. Then we 
used the Galactic model with the same atmospheric parameters 
but scaled the metallicity in the {\sc synplot} 
spectrum synthesis calculation in order to mimic that of the SMC. 
The resulting spectra were very similar with the summed equivalent 
widths differing by 3\%. Hence we do not expect that this 
simplification (that has been widely used elsewhere -- see, for example
Kilian et al. \citealp{Kil94}, Korn et al. \citealp{Kor00}, McErlean et al.
\citealp{McE99}), will be a source of significant uncertainty.
However it is possible that this approach could lead to 
uncertainties that depended on the wavelength region observed as 
the blanketing will depend on the iron abundance adopted. In turn this 
could affect the ionization structure and thereby the behaviour of
absorption lines formed in different parts of the stellar atmosphere.

As discussed in Sect. \ref{Obs_red}, it is difficult to distinguish
continuum regions due to the richness of the ultra-violet absorption line 
spectrum. Hence we have not attempted any continuum normalisation but 
have instead worked directly with the flux calibrated spectra. For the 
observed spectra, we must first correct for the effect of interstellar 
extinction. We adopted values of the colour excess
(summarized in Table \ref{Obs_data}) for the Bridge targets
from Demers \& Battinelli (\citealp{Dem98}), for LH\,10-3270 from Parker et
al. (\citealp{Par92}) and for AV\,304 from Azzopardi \& Vigneau 
(\citealp{Azz82}).
There is evidence that the extinction law for the Magellanic Clouds 
may differ from that for the Galaxy (see, for example, Seaton
\citealp{Sea79}, Howarth \citealp{How83}). However given the low
reddenings of our targets the use of different laws was found to have
only a small effect on the absolute fluxes deduced. More importantly,
because of the limited  coverage  of the observed spectra, the relative
flux distributions, which are important  for our methodology as
discussed below, remain effectively unchanged when corrected for
extinction. Hence uncertainties in these corrections are unlikely to
seriously affect our analysis. We then rebinned our unreddened spectra
to a wavelength interval of 0.1\AA\ and to a wavelength  range of
1890--1970\AA\ for the STIS spectra and 1890--1928\AA\ for the GHRS
spectra.

To compare our observed and theoretical spectra, we adopted the
following  methodology. The theoretical spectra were first convolved
with a Gaussian broadening function to allow for intrumental broadening 
and with a rotational broadening function with the projected rotational
velocities, \vsini, being summarized in Table \ref{Atm_par}. These were
taken from Hambly et al.\  (\citealp{Ham94}) for DI\,1388, Rolleston et
al.\ (\citealp{Rol99}) for DI\,1239 and DGIK\,975, Rolleston et al.\
(\citealp{Rol02}) for LH\,10-3270 and Rolleston et al.\
(\citealp{Rol03}) for AV\,304. Recent studies (see, for example,
Dufton et al. \citealp{Duf06}, Sim\'on-D\'iaz and Herrero
\citealp{Sim07}) have shown that macroturbulence can also be important
in B-type supergiants. The estimated gravities of our targets imply that they
are either near to the main sequence or giants and hence that macroturbulence
will not be significant. Additionally we use these projected rotationally 
velocity estimates to allow for the overall broadening of the line spectra and
to first order its physical origin is not important. 
The theoretical spectra were then
rebinned to the same wavelength region and interval as those of the 
observed spectra and scaled so that they had the same mean flux as the
observed spectra, thereby making the spectra  comparable without any
need for rectification or normalisation. The sum of the squares of the
differences between observation and theory (S) was then computed  and
the metallicity varied until this was minimised. These metallicity
estimates (designated as $\Delta [\frac{Fe}{H}]$ because our
ultra-violet  spectra are dominated by iron group absoption lines) are
summarized in Table \ref{Fe_abund}. 

\begin{figure*}
\includegraphics[angle=270,width=6.0in]{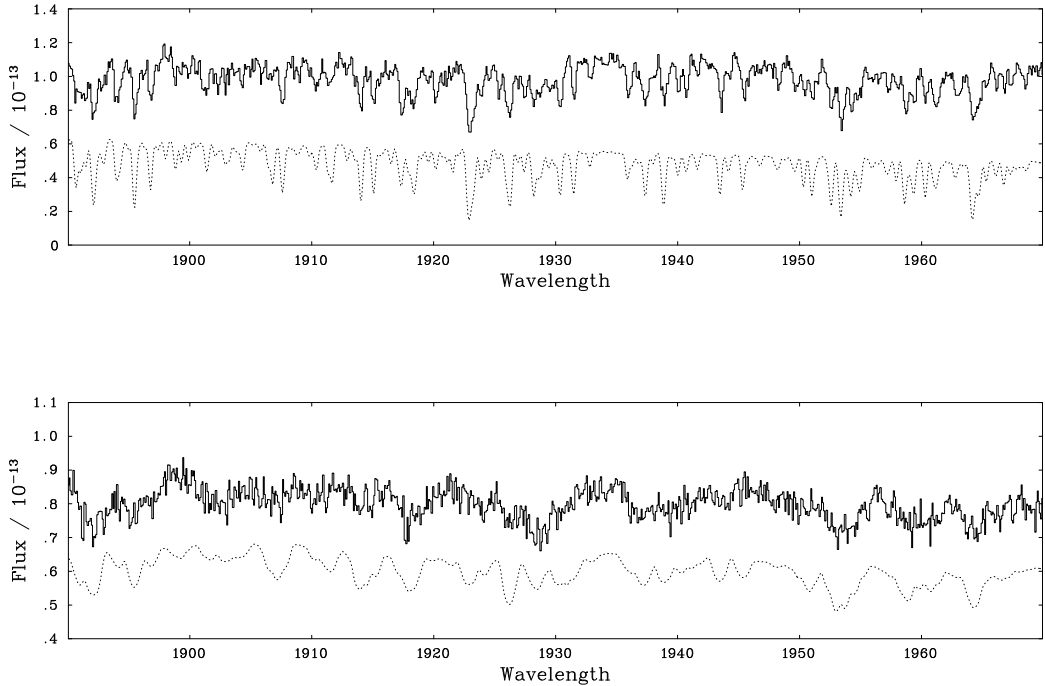}
\caption{Comparison of observed and theoretical spectra for two of our targets.
The upper panel shows the observed (solid line) and the theoretical 
(dotted line) spectra for LH10\,3270 and a metallicity of -0.9 dex 
(corresponding to the best fit between observation and theory)
relative to that of the Galaxy. To aid comparison the theoretical
spectrum has been shifted downwards on the y-axis. The lower panel shows
the corresonding spectra for DGIK\,975, a metallicity of -1.7 dex now being
adopted for the theoretical calculations}
\label{1029_fig1}
\end{figure*}

\begin{figure}
\includegraphics[angle=0,width=3.0in]{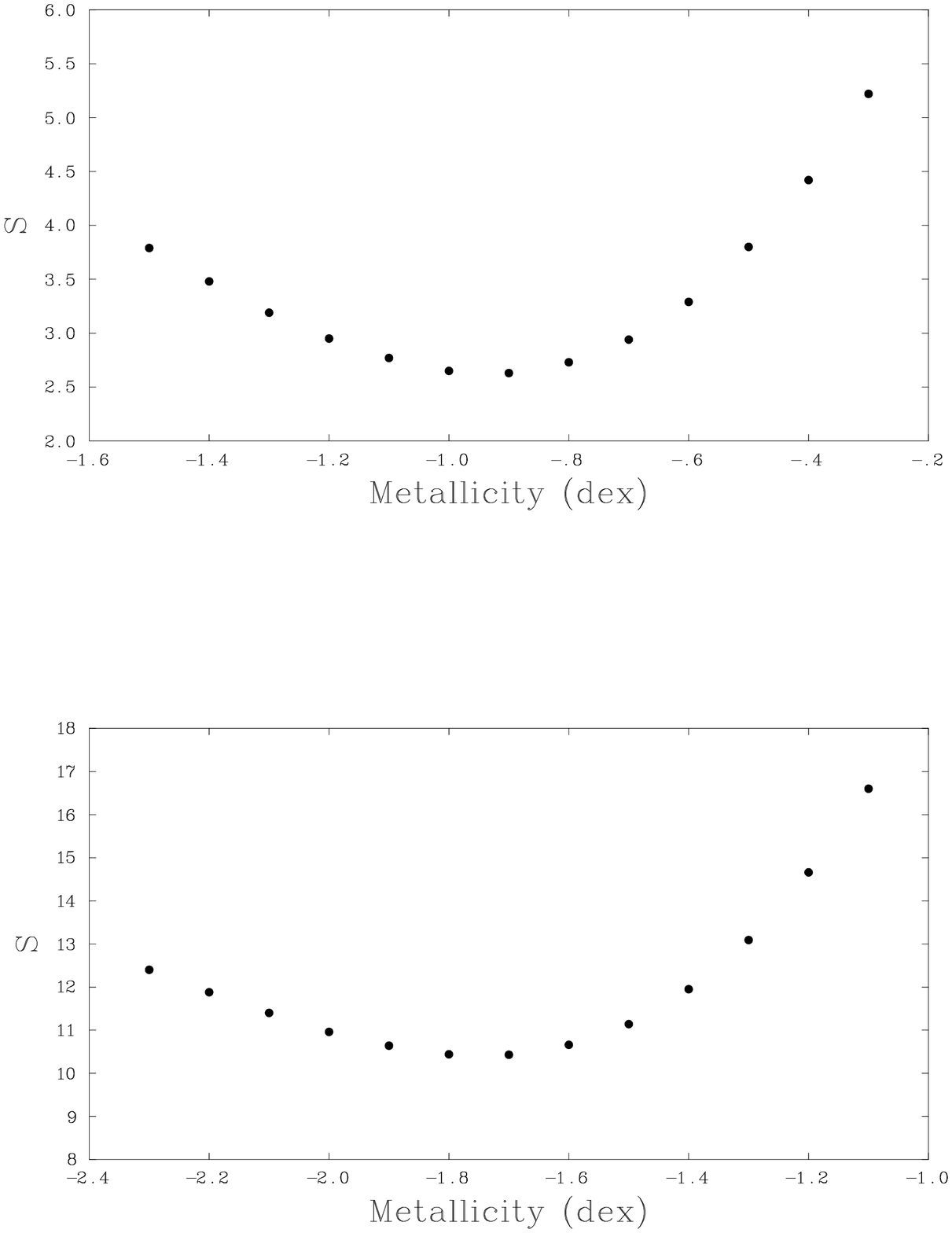}
\caption{Estimation of the metallicity for two of our targets. The upper panel
shows the goodness-of-fit, S (with units of 
10$^{-28}$ erg$^2$ cm$^{-4}$s$^{-2}$\AA$^{-2}$) 
between the observed and theoretical spectra
for LH10\,3270 plotted against the metallicity adopted for the theoretical
calculations. The lower panel shows the analagous results for DGIK\,975}
\label{1029_fig2}
\end{figure}

This approach is illustrated in Figs.  \ref{1029_fig1} and \ref{1029_fig2}
for  the LMC target, LH\,10-3270 and the Bridge target, DGIK\,975
chosen to cover a range of metallicities and projected rotational
velocities. In the former the observed and best fitting theoretical
spectra are shown, whilst in the latter, the value of S is plotted against
metallicity and   minima are found at metallicities (relative to
Galactic) of -0.9 and -1.7 dex respectively. In general the
theoretical spectra reproduce that observed but agreement for
specific features is variable. The absoption line spectrum for
LH\,10-3270 is relatively well observed due to both its low projected
rotational velocity and the relatively high metallicity of the LMC. The
theoretical spectrum is in excellent agreement with the majority of
absorption features well matched. This in turn provides indirect
support that our uniform scaling  of all the individual element
abundances is acceptable.

For DGIK\,975, the situation is less satisfactory with the match to 
individual features being far less convincing, which may be manifested 
in the broad minimum found in the goodness-of-fit shown in 
Fig.\ \ref{1029_fig2}. Given the better agreement found for 
LH\,10-3270, we ascribe these problems principally to the observed 
absorption lines being intrinsically weak and broadened by the significant 
stellar rotation. However part of these discrepancies may also be due 
to a combination of inadequacies in the atomic data (and in particular 
the oscillator strengths) and the neglect of non-LTE effects for the 
species that dominate the absorption in this spectral region.

The spectra line data used in {\sc synspec} is mainly taken from the
linelists of Kurucz (see, for example, \citealp{Kur88,Kur91}). These have been 
updated using more recent results taken from the NIST database as discussed
in for example Lanz and Hubeny (\citealp{Lan03, Lan07}), with further
information being available from the {\sc tlusty/synspec} webpage. All the
energy levels are experimental and as such the corresponding
wavelengths of the transistions should be reliable. This was confirmed by
comparing the our wavelengths for a sample of Fe {\sc iii} features with those
determined experimentally (see, Sugar and Corliss, \citealp{Sug85} and 
references; Ekberg, \citealp{Ekb93}
therein) where discrepancies were typically less than 0.01\AA. This would
correspond to a velocity uncertainty of approximately 1 km s$^{-1}$, which is 
far smaller than the intrinsic width of the line due to the rotational 
broadening. Hence we do not believe that uncertainties in the wavelengths 
will be a significant source of error. 
    
In the case of the oscillator strengths, a number of independent
calculations exist apart from those of Kurucz (\citealp{Kur91}), including
Toner and Hibbert (\citealp{Ton05}), Ekberg (\citealp{Ekb93}), Fawcett
(\citealp{Faw89}), Nahar and Pradhan (\citealp{Nah96}). Raassen and Uylings
have also provided extensive listings via a web-site
(http://www.science.uva.nl/pub/orth/iron/FeIII) as part of the
{\sc ferrum} consortium (Johansson et al., \citealp{Joh00}). As discussed
by Toner and Hibbert, discrepancies are typically 40\%, which for a single
feature would lead to an absolute abundance error of approximately 0.2 dex.
However as discussed by Thompson et al. (\citealp{Tho07}), for the stronger
lines the discrepancies are smaller, being typically 10\% or less.
Additionally as we fit a large number of features simultaneously, the 
corresponding abundance error would be smaller if these discrepancies were 
random. The
errors in differential abundances should also be smaller as errors due to
uncertainties in the oscillator strengths should to first order cancel. 
As even for the best modern calculations, there remains differences 
between the theoretical oscillator strength estimates and as these new 
results cover only a subset of the lines incorporated in our linelists, 
we have decided not to attempt to update our linelist.
Given the above discussion the abundance estimates (listed in Table 
\ref{Fe_abund}) must be treated with caution. However variations in these 
estimates should be more reliable as they are based on a consistent and 
objective approach.

The above methodology implicitly assumes that the flux calibration of the
observed spectrum has been reliably performed. For example there could remain
a low frequency ripple and to investigate the consequence of this we have
artificially added a sine wave component with an amplitude of 1\% and a period
equal to the wavelength range of our data for the target, LH\,3270. We then
repeated the analysis and found the same abundance estimate (after rounding to
the nearest 0.1 dex). Clearly larger errors would map onto larger 
uncertainities in the abundances but 
we would expect that any systematic errors in the calibration would similarly
affect all our abundance estimates and hence relative values should be more
reliable.

One significant discrepancy in the comparison shown in Fig. 
\ref{1029_fig1} is that the theoretical spectra 
predicts a relatively strong feature at approximately 1906.8\AA. This is 
not apparant in the observed spectra for LH\,3270 and also may be absent from 
that for DGIK\,975 although in this case the comparison is complicated by the
greater amount of rotational broadening and the intrinsic weakness of the
spectrum. Investigation of the theoretical spectra indicates that this feature 
arises principally from an Fe {\sc iii} line, which has been incorporated
with and oscillator strength 0.78. For this transition, Raassen and Uylings
derive a smaller value of 0.41, which may at least partially account for the
discrepancy. As a test, we have repeated the fitting procedure for 
LH10\,3270 excluding this region. The goodness-of-fit was indistinquishable
from that shown in Fig. \ref{1029_fig2} and lead to the same estimate of the
iron abundance as listed in Table \ref{Fe_abund}.

For DI\,1388, the wavelength coverage of its GHRS spectrum is approximately 
half of that of the STIS observations and this leads to its metallicity
estimate not being strictly comparable with those for the other targets. 
Additionally the S/N ratio of the GHRS spectrum is relatively low, whilst 
DI\,1388 has the highest \vsini\ estimate. Hence the metallicity estimate 
for this star is particularly uncertain, although we note that from visual
inspection the agreement between observation and theory appears poor for
values of the metallicity, $\Delta [\frac{Fe}{H}]> -1.3$\,dex.

Inspection of Table \ref{Atm_par} indicates that the two SMC and LMC targets
have smaller projected rotational velocities and hence their spectral lines will
be narrower. In turn this could affect the comparison between iron abundances
deduced in these targets and those in the Bridge. We have investigated this as
follows. For these two targets we have rotationally broadened their observed
spectra to a value of \vsini = 120 \kms, which is typical of that observed in
our Bridge targets. We have then added additional Poisson noise to obtain a
S/N ratio of 30 again typical of the Bridge targets. We have then repeated the
analysis of these modified spectra to obtain estimates of the iron abundances.
Although there are small changes in the position of the minimum obtained in
the goodness-of-fit, after rounding the abundance estiamtes to the nearest 0.1
dex, we obtain the values listed in Table \ref{Fe_abund}. Hence we do not
believe that the differences in broadening in the observed spectra should affect
estimates of the abundance differences between the Bridge and the Cloud
targets.

\begin{center}
\begin{table*}
\caption[]{Estimated iron abundances relative to that of our Galaxy (i.e. 
$\Delta [\frac{Fe}{H}]=0$\ dex would correspond to a solar metallicity
of 7.5 dex - Grevesse and Sauval \citealp{Gre98}) for our targets
deduced from the HST ultra-violet spectra. Mean light metal abundance 
estimates previously deduced for these  targets (designated $\Delta
[\frac{X}{H}]$ - see text for details) are also summarized.  Other
estimates of the iron abundances for the LMC and SMC relative to our
Galaxy are tabulated. These are from B-type supergiants by Trundle et
al. \cite{Tru07},  A-type supergiants by Venn (\citealp{Ven99}), 
F-type supergiants by  Andrievsky et al. (\citealp{And01}) and Hill
(\citealp{Hil97}). The final column lists abundance studies of
late-type stars, viz. Cepheids in the LMC (Luck et al.\
\citealp{Luc98}) and K-type  supergiants in the SMC (Gonzalez and
Wallerstein \citealp{Gon99}).
}

\begin{tabular}{lccccccccc}\hline
Star  & Region & $\Delta [\frac{Fe}{H}]$ & $\Delta [\frac{X}{H}]$ &
\multicolumn{4}{c}{$\Delta [\frac{Fe}{H}]$}
\\
& & & & B-type & A-type & F-type & late-type
\\
 & & dex & dex & dex & dex & dex & dex
\\
\hline
LH\,10-3270  & LMC    & -0.9 & -0.4    & -0.3  & --   & -0.4 & -0.3
\\
AV\,304      & SMC    & -1.2 & -0.6    & -0.6  & -0.8 & -0.7 & -0.6
\\              
DGIK\,975    & Bridge & -1.7 & -1.2    & --    & --   & --   & --
\\              
DI\,1239     & Bridge & -1.6 & -1.1    & --    & --   & --   & --
\\               
DI\,1388     & Bridge & -1.9 & $<$-0.9 & --    & --   & --   & --
\\
\hline
\end{tabular}
\label{Fe_abund}
\end{table*}
\end{center}

We have estimated the uncertainties in our Fe abundance estimates
arising  from uncertainties in the adopted atmospheric parameters. We
assumed errors of $\pm$1\,000 K in \teff, $\pm$0.2 dex in \logg\ and $\pm$3
\kms   in \vt, which is consistent with the error estimates discussed
in the analyses  summarized in Table \ref{Atm_par}. We have used a
relatively simple procedure by first calculating the total equivalent
width for the  ultra-violet spectra region that we have analysed for
our adopted   stellar atmospheric parameters and our estimated iron
abundances  (see Tables \ref{Atm_par} and \ref{Fe_abund}
respectively). We then increased in turn each of the atmospheric
parameters by its  uncertainty and changed the iron abundance until we
recovered the same total equivalent width. Tests comparing this
methodology with that of changing each atmospheric parameter and then
recalculating the goodness-of-fit between theory and observation for
different Fe abundances showed that the two methods yielded similar
error estimates. Our error estimates are summarized in 
Table\,\ref{Fe_error}, with values for a similar decrease in the
atmospheric parameters being of similar magnitude but with an opposite
sign. In general, the errors arising from uncertainties in the surface
gravity appear not to be significant. Those for the microturbulence
depend on the strength of the Fe  absorption lines with those for the
effective temperature depending on how far the effective temperature is
from that (approximately 18\,000 K)  corresponding to the maximum in
the strength of the Fe spectrum. Assuming that these errors are
uncorrelated, they can be added in quadrature and these error estimates
are shown the final column of the  Table\,\ref{Fe_error}. Adopting
different errors estimates for  the atmospheric parameters or allowing
for the correlation between  the effective temperature and gravity
estimates would lead to different  estimates but we believe that a
typical uncertainty of $\pm$0.2 dex implied by Table\,\ref{Fe_error} to
be reasonable.

We have also attempted to use the value of the $\chi^{2}$\ associated
with our comparison of the theoretical and observed spectra to estimate
the uncertainty due for example to observational errors. As discussed
by Avni (\citealp{Avn76}), a change of 1.0 and 2.7 in the $\chi^{2}$\
parameter should correspond to the 68 and 90\% limits in our
uncertainty. However utilisation of this procedure  led to unrealistic
small error estimates of less than 0.1 dex for both cases. This  may
reflect that some of our data points (e.g. those near the continuum) do
not represent signal but tests attempting to exclude such regions again
led to low error estimates. Additionally the small implied errors may 
be due to stochastic noise not being the dominant source of uncertainty.
For example other systematic errors in the observational data or from the
physical assumptions (including the adopted atomic data) may be significant.

Moehler et al. (\citealp{Moe98}) in their
analysis of the equivalent widths of individual iron features in the
ultra-violet spectrum of PAGB stars estimated errors due to
observational uncertainties of 0.05 to 0.1 dex. From inspection by eye
of the quality of our fits for different Fe abundances an error of 
approximately
$\pm$0.2 dex would appear reasonable which is rather larger than the
errors quoted by Moehler et al.  (\citealp{Moe98}). However adopting a
typical random error of $\pm$0.2 dex and combining in quadrature with
those due to uncertainties in the  atmospheric parameters leads to an
estimated error of typically $\pm$0.3 dex. We note that this estimate
does not include systematic errors from, for example, the assumption of
local thermodynamic equilibrium in the radiative transfer calculetion
and we return to this point in Sect.\ \ref{Discussion}.

\begin{center}
\begin{table}
\caption[]{Uncertainties in our estimated iron abundances due to 
uncertainties in the atmospheric parameters. For each star, the
effective temperature, logarithmic gravity and microturbulence have
been increased by 1\,000 K, 0.2 dex and 3 \kms respectively and the 
change of iron abundance required to obtain the best agreement 
between theory and observation was estimated; see text for further 
details. Also shown is the estimated errors assuming that these
errors can be added in quadrature.
}
\begin{center}\begin{tabular}{lccccccccc}\hline
Star  & $\Delta T_{\rm eff}$ & $\Delta$ \logg & $\Delta$\vt & Quadrature
\\
      & +1\,000 K            & +0.2 dex       & +3 \kms 
\\
\hline
LH\,10-3270  & +0.18 & -0.06 & -0.09 & $\pm$0.21
\\
AV\,304      & +0.17 & -0.03 & -0.16 & $\pm$0.24
\\              
DGIK\,975    & -0.06 & +0.03 & -0.12 & $\pm$0.14
\\              
DI\,1239     & +0.09 & +0.03 & -0.12 & $\pm$0.15
\\               
DI\,1388     & +0.18 & -0.06 & -0.02 & $\pm$0.19
\\
\hline
\end{tabular}\end{center}
\label{Fe_error}
\end{table}
\end{center}

\section{Discussion}
\label{Discussion}
In Table \ref{Fe_abund}, we also summarize for our targets, the mean 
differential abundance  (designated $\Delta [\frac{X}{H}]$) for all 
reliably observed elements from carbon to sulphur with the exception of
nitrogen. This is because analysis of H {\sc ii} region  (see, for example,
Kurt et al. \citealp{Kur99}; Peimbert et al.  \citealp{Pei00}; Testor
\citealp{Tes01}; Tsamis et al. \citealp{Tsa03})  and stellar spectra (see,
for example, Korn et al. \citealp{Kor02, Kor05}) indicate that nitrogen has
anomalously low differential abundances in the  Magellanic system. The
sources of these {\em light element} abundances are:  LH\,10-3270
(Rolleston et al.\ \citealp{Rol02}); AV\,304 (Rolleston et al.\
\citealp{Rol03}; Hunter et al. \citealp{Hun05}); DKIK\,975 and DI\,1239
(Rolleston et al.\ \citealp{Rol99}) and DI\,1388 (Hambly et al.
\citealp{Ham94}). In this Table, we also summarize {\em iron} abundances
estimated for some representative analyses of other stellar targets: SMC
and LMC B-type supergiants (Trundle et al.  \citealp{Tru07}); SMC A-type
supergiants (Venn \citealp{Ven99}); F-type supergiants (Andrievsky et al.
\citealp{And01}; Hill  \citealp{Hil97}); LMC Cepheids (Luck et al.\
\citealp{Luc98}) and  SMC K-type supergiants (Gonzalez \& Wallerstein
\citealp{Gon99}).

A comparison of our iron abundances deduced from ultra-violet spectra  with
those obtained for other stellar targets shows that the former are lower by
0.5-0.6 dex for both the SMC and  LMC. This would appear to be the case
even when we compare the analyses of isolated {\it optical} Fe {\sc iii}
lines of B-type stars  with our ultra-violet spectra (which are again
dominated by Fe {\sc iii} lines), although  unfortunately there are no
stars in common between the analyses. Studies of the ultra-violet spectra
of other hot stars have also found relatively low iron abundances, as
discussed by Thompson et al. (\citealp{Tho07}). For example, Grigsby et al.
(\citealp{Gri96}) analysed  the iron lines between 1800 and 2500 \AA\ in
the spectrum of the  B3 V star $\iota$\,Her. They found an underabundance
of -0.47 dex  relative to the sun assuming a zero microturbulence. As
pointed out  by Gribsby et al., an increase in the microturbulence would
decrease  their estimates and hence the actual underabundance estimated
from the ultra-violet spectrum could be larger. From their results,  we can
deduce an abundance estimate from the Fe {\sc iii} lines that lie in the
wavelength region (1888 to 1978 \AA) that we have  considered. This again
implies that iron is underabundant  by 0.4 dex or more (depending on
the microturbulence adopted)  compared with the solar value (Meyer
\citealp{Mey85}; Anders and Grevesse \citealp{And89}; Grevesse and Sauval
\citealp{Gre98}). By contrast, analyses  of the optical Fe {\sc iii}
spectrum in $\iota$\,Her (Kodeira and Scholz  \citealp{Kod70}; Peters and
Aller \citealp{Pet70}; Peters and Polidan  \citealp{Pet85}; Pintado and
Adelman \citealp{Pin93}) have led to values  that are close to solar. 

Hot post-asymptotic-giant branch (PAGB) stars have also been found with 
relatively low iron abundances as discussed by Moehler \cite{Moe01}. 
Napiwotzki et al. (\citealp{Nap94}) analysed optical and ultra-violet
spectra of the field star BD+33$\deg$2642 and found underabundances
(relative to solar) of  approximately  1.0 dex for C, N, O, Mg, Si but a
larger underabundance of 2.0 dex for Fe. The latter was found from analysis
of Fe {\sc iii} lines in the same spectral  region as considered here.
Napiwotzki et al. showed that such an abundance pattern could be understood
by a seperation of gas and dust as proposed for  cooler PAGB stars (see,
for example, Mathis and Lamers \citealp{Mat92}). Moehler et al.
(\citealp{Moe98}) analysed ultra-violet and optical spectra of two PAGB
stars, ROA\,5701 and Barnard 29, in globular clusters with the main aim of
determining their iron abundance. From the ultra-violet spectra, they
determined iron abundances that for Barnard 29 was 0.5 dex lower than  that
found for the cooler giant stars in M\,13 and for ROA\,5701 was lower  than
the range of abundances found in $\omega$\,Cen. Moehler et al. again
interpreted these observations as due to a gas/dust separation in the 
atmospheres of the PAGB stars. Recently Thompson et al. 
(\citealp{Tho07}) have reanalysed the spectroscopy for these stars. From
the ultraviolet spectra they find iron abundances using the spectra
synthesis technique used here that are similar to those estimated by 
Moehler et al. However they also tentatively identified Fe {\sc iii} lines
in their optical spectra. These implied higher iron abundances that agree
well with those found from the red giants in M\,13 and with the range
of iron abundances found from the red giants in $\omega$\,Cen (Origlia et
al.\ \citealp{Ori03}).

In summary analyses of the ultraviolet Fe {\sc iii} spectrum of B-type 
stars around 1900 \AA\ have consistently yielded relatively low iron
abundances. For the highly evolved objects, this may reflect real physical
effects but  in others there appears to be differences between abundance
estimates  from the optical and ultra-violet spectra and for the younger
massive stars  inconsistencies with the iron abundance that would be
expected. Indeed our results for the SMC and LMC targets and those
discussed above for  $\iota$ Her would imply that a systematic increase in
the iron abundances  deduced from the ultra-violet spectra of approximately
0.5 dex would appear appropriate. The cause of this systematic offset could
be in the theoretical  modelling and in particular in the use of an LTE
approximation in the spectral synthesis or in the adopted oscillator
strengths. Additionally if the ionic linelists are incomplete, the presence
of a large number of weak unresolved lines would act as a pseudo-continuous
opacity. In turn this could lead to the strengths of lines or blends in our
theoretical spectra being too strong, thereby leading to our abundance
estimates being too small.

Given the uncertainties, in the absolute iron abundances deduced from the HST
spectra, we have instead chosen to take a differential approach. For
example,  we find a difference between in the iron abundance of our SMC and
LMC target  of -0.3 dex which is in excellent agreement with those found
from the optical spectra of BAF-type and late-type stars in the Magellanic
Clouds (see Table \ref{Fe_abund}). More importantly all three Magellanic
Bridge targets  appear to have very low iron abundances and in particularly
values  that are lower than the SMC iron abundance by between -0.4 to -0.7
dex.  These differences are larger than the error estimates discussed in
Sect.  \ref{spec_anal} and although it is possible that these have been 
underestimated, the similarity in the iron underabundance found for all 
three Bridge targets is striking. Additionally as we have adopted a
differential approach, any systematic errors should have been minimised.

Such an iron underabundance is consistent  with those for the lighter metals
(listed in Table  \ref{Fe_abund}) and with the analysis discussed in
Rolleston et al.  (\citealp{Rol99}) who estimated a mean light metal
underabundance for the Bridge material of  -0.60$\pm$0.18 dex with respect to
the SMC. Additionally Lehner et al. (\citealp{Leh01}), from an analysis of the
interstellar absorption lines  seen in the spectra of a background object,
also found evidence for low  metal abundances in the Magellanic Bridge gas.
Therefore in the discussion  that follows, we will assume that the material
in the Magellanic Bridge is underabundant in metals by factors of
approximately -0.5 and -0.8 dex compared with the SMC and LMC  respectively.

As discussed in the introduction, both observational (e.g. Irwin et al.
\citealp{Irw85,Irw90};  Demers \&\ Irwin \citealp{DIr91}) and theoretical
(e.g. Gardiner and Noguchi  \citealp{Gar96}; Sawa et al. \citealp{Saw99})
studies imply that the Bridge was formed during a close
encounter between the Magellanic Clouds approximately 200 million years ago.
This scenario has been supported by more recent investigations, with for
example Nishiyama et al.  (\citealp{Nis07}) tentatively identifying pre-main
sequence Herbig Ae/Be stars in the Magellanic Bridge. Additionally the
recent  photometric investigation of Harris (\citealp{Har07}) 
failed to identify an older stellar population implying that the
material stripped from the Magellanic Clouds was effectively a pure gas.

Of particular relevance to the metal abundances found
in the Magellanic Bridge are the recent numerical simulations of Bekki and
Chiba (\citealp{Bek05,Bek07}). These follow the evolution and star
formation history of the Magellanic System over the last 800 million
years using chemodynamical simulations with both Clouds being modeled
as self-gravitating barred and disk systems. A mean metal abundance
for the Bridge (relative to our Galaxy) of -0.8 dex is predicted,
compared with -0.6 dex adopted for the cental regions of the SMC. This
underabundance arises from the Bridge material primarily originating
from the outer regions of the SMC with the simulations incorporating a
negative metallicity gradient. As pointed out by Bekki and
Chiba (\citealp{Bek07}), this is qualitatively consistent with the
Bridge light element underabundances found by Rolleston et al. 
(\citealp{Rol99}) and summarized in Table \ref{Fe_abund}. However as
for our iron abundance estimates of the three bridge stars, there
remains a significant quantitative discrepancy of approximately 0.3
dex.

Bekki and Chiba (\citealp{Bek07}) predicted a range of metallicity in
Bridge stars with a mean value of -0.8 dex but with approximately
three percent of stars having values $\leq -0.9$ dex. Our three targets are
spatially well separated across the Magellanic Bridge (see Figure 3 
of Rolleston et al. \citealp{Rol99}) and hence it would be highly 
improbable that they could all be from the lower metallicity part 
of this distribution. We therefore conclude that a plausible
explanation for the very low light metal and iron abundances found 
in the Magellanic Bridge is that the corresponding material was
stripped from the outer regions of the SMC approximately 200 million
years ago. However observational abundance estimates and theoretical 
predictions still differ by 0.3 dex. 

\section{Conclusions}
\label{Conclusions}
We have presented HST ultra-violet spectroscopy of five targets in the 
Magellanic Clouds and Bridge. These data have been analysed using 
grids of non-LTE models atmospheres with the spectral synthesis being
predominantly undertaken in an LTE approximation to deduce
metallicities (which will be dominated by the iron abundances). 
The principal results can be summarized as follows:
\begin{enumerate}
\item Iron abundances estimated from the ultra-violet spectra of SMC
and LMC targets are typically -0.5 dex lower than those derived by 
other methods. This is consistent with results found from previous 
studies of Galactic objects.
\item The iron abundance estimates for the three bridge targets are
-0.5 and -0.8 dex lower than those for the SMC and LMC respectively.
This is consistent with the underabundance in light metals of -1.1 dex
found for the Bridge material with respect to our Galaxy.
\item These metal abundances are in qualitative agreement with
numerical simulations which imply that most of the Bridge material has
been stripped from the outer layers of the SMC. However there remain
significant quantitative differences between the observed estimates
and those predicted by the simulations.
\end{enumerate} 
\section*{Acknowledgments}
We acknowledge financial support from PPARC (now STFC) and are
grateful for useful discussions with Ian Howarth and Alex de Koter.
HMAT acknowledges financial support from the Northern Ireland Department 
for Education and Learning (DEL).

\bsp

\label{lastpage}

\end{document}